\documentstyle[aps,tighten,preprint,floats,12pt,epsf]{revtex}
\newcommand{\li}{\mathop{{\mbox{Li}}_4}\nolimits}
\newcommand{\aagt}{\,\rlap{\lower 3.5 pt \hbox{$\mathchar \sim$}} \raise 1pt
 \hbox {$>$}\,}
\newcommand{\aalt}{\,\rlap{\lower 3.5 pt \hbox{$\mathchar \sim$}} \raise 1pt
 \hbox {$<$}\,}

\begin{document}

\preprint{
\hbox to \hsize{
\hfill$\vtop{   \hbox{MPI/PhT/98--73}
                \hbox{hep-ph/9809507}
                \hbox{September 1998}
                \hbox{}
                \hbox{}}$}
}
\title{Higgs Decay into Gluons up to ${\cal O}(\alpha_s^3 G_F m_t^2)$}
\author{Matthias Steinhauser\thanks{Address after Oct. 1.:
Institut f\"ur Theoretische Physik, Universit\"at Bern, Sidlerstrasse 5,
CH-3012 Berne, Switzerland}
}
\address{Max-Planck-Institut f\"ur Physik,
    (Werner-Heisenberg-Institut),\\ D-80805 Munich, Germany}
\maketitle

\thispagestyle{empty}

\begin{abstract}
The decay of the Standard Model Higgs boson in the
intermediate-mass range into
gluons is considered where special emphasis is put on the
influence of the
leading electroweak corrections proportional to $G_F m_t^2$.
An effective Lagrangian approach is
used where the top quark is integrated out. 
The evaluation of the coefficient function is
performed using two different methods.
The first one is concerned with the direct evaluation of the vertex
diagrams and the second method is based on a low-energy theorem.
In a first step the tools needed for the computation are provided
namely the renormalization constants of the QCD Lagrangian are computed up
to ${\cal O}(\alpha_s^2 G_F m_t^2)$. Also the decoupling constants for
the strong coupling constant $\alpha_s$ and the light quark masses
relating the quantities of the full theory to the corresponding
quantities of the effective one are evaluated up to order
$\alpha_s^2 G_F m_t^2$.

\medskip

\end{abstract}

\newpage
\setcounter{page}{1}


\section{Introduction}

The Higgs boson is the only not yet discovered particle of the
Standard Model of elementary particle physics.
Up to now only lower bounds on its mass of 
$M_H\aagt89.8$~GeV~\cite{McNVan}
could be derived from the lack of observation at LEP~1
and LEP~2.
Through the virtual presence of the Higgs boson in loop diagrams it is also
possible to set indirect limits on $M_H$ with the help of the
precision data collected at LEP, SLC and Tevatron.
Currently they read $M_H= 84^{+91}_{-51}$~GeV with an upper limit
of $280$~GeV at 95~\% C.L.~\cite{KarVan}.
These numbers suggest that a Higgs boson in the so-called intermediate-mass 
range, i.e. $M_H\aalt2M_W$ where $M_W$ is the mass of the $W$ boson, 
is an attractive candidate. In this paper we will therefore consider
such a Higgs boson and compute corrections to its gluonic decay rate.

Comprehensive reviews concerning the properties of the Higgs boson
are given in~\cite{Kni94,Spi98}.
The dominant decay mode of an intermediate-mass Higgs boson
is the one into bottom quarks. The QCD corrections are known up to
${\cal O}(\alpha_s^3)$~\cite{DreHik90,GorKatLarSur90,Che97CheSte97}.
Concerning the electroweak theory, the full one-loop corrections are
available~\cite{Kni92DabHol92}. For the mixed electroweak/QCD corrections only 
the leading terms of order 
$\alpha_s G_F m_t^2$~\cite{KwiSte94KniSpi94} and 
$\alpha_s^2 G_F m_t^2$~\cite{KniSte95,CheKniSte97hbb}
are evaluated at the two- and three-loop level, respectively.
An other important decay mode is the one into gluons. However, this process
is suppressed as compared to the fermionic channel
as at lowest order it is mediated via a
quark loop. It turns out that the leading order QCD corrections 
are quite large and amount to roughly 
$66$~\%~\cite{InaKubOka83,DjoSpiZer91}.
Recently also the
next-to-leading terms were evaluated which give a further 
enhancement of roughly $20$~\% and thus increase the confidence to use
perturbative QCD~\cite{CheKniSte97hgg}.
In~\cite{DjoGam94}
the leading electroweak corrections of order $G_F m_t^2$
were evaluated. In turned out that cancellations between different
contributions take place and the final result is quite small.
Nevertheless it is interesting to add an extra gluon and to evaluate the
three-loop corrections of order $\alpha_s G_F m_t^2$, also in order to
observe the behaviour of the perturbative series.
Thus the aim of this paper is to consider next to QCD corrections
also those terms arising from the electroweak theory which are 
enhanced by the top quark mass and proportional to $G_F m_t^2$.

For a Higgs boson in the intermediate mass range it makes sense to 
consider the limit $M_H^2\ll m_t^2$ and to construct in a first
step an effective Lagrangian where the top quark is integrated
out. Then the main task for the computation of the leading electroweak
corrections is the evaluation of the effective coupling of the
Higgs boson to gluons usually called $C_1$. The operators appearing
in the effective Lagrangian are only defined in the effective
theory and therefore receive no corrections involving the top quark.
Note that
$C_1$ enters not only into the decay rate but constitutes also a building
block for the gluon fusion process which will be the dominant
mechanism for the production of the Standard Model Higgs boson
at the CERN Large Hadron Collider. 
The effective coupling to quarks is usually denoted by $C_2$
and was computed up to three loops in~\cite{CheKniSte97hbb}.

It is well-known that the Appelquist-Carazzone decoupling 
theorem~\cite{AppCar75} does not hold true in its naive sense if
a renormalization scheme based on minimal subtraction is used.
This means that the contribution from a heavy quark $h$ with mass $m_h$
to a Green function of gluons and light quarks does in general not
show the expected $1/m_h$ suppression.
The standard solution to this problem is to do the decoupling
``by hand'' and to construct an effective theory where the
heavy quark is integrated out. The quantities relating the parameters,
respectively, fields of the full theory to the corresponding
quantities in the effective one are called decoupling constants.
In~\cite{CheKniSte98dec} it was shown that there is a tight connection
between the (renormalized) coefficient functions, $C_1$ and $C_2$, and
$\zeta_g$, respectively, $\zeta_m$ which perform the decoupling for the strong
coupling constant $\alpha_s$ and the light quark masses $m_q$.
Concerning pure QCD, $\zeta_g$ and $\zeta_m$ are known
up to the three-loop level~\cite{Wei80,BerWet82,LarRitVer95,CheKniSte98dec}.
In this paper the leading electroweak corrections are considered and
terms up to ${\cal O}(\alpha_s^2 G_F m_t^2)$ are evaluated.
The two-loop terms of order $\alpha_s G_F m_t^2$ can be found
in~\cite{CheKniSte97hbb}.

The organization of the paper is as follows:
In the next Section the notation is fixed and 
the theoretical framework developed in
previous papers is reviewed.
Section~\ref{secren} is concerned with the computation of the
renormalization constants for the QCD parameters and fields up
to order $\alpha_s^2 G_F m_t^2$.
Although only the ones for the coupling constant $\alpha_s$ and the
light quark masses are needed we provide all renormalization constants
of the QCD Lagrangian up to this order. 
In Section~\ref{secdec}
$\zeta_g$ and $\zeta_m$ are computed
up to order $\alpha_s^2 G_F m_t^2$. 
Afterwards, in Section~\ref{sechad}, they are used in order to
compute the coefficient functions $C_1$ and $C_2$. The result
for $C_1$ is
compared with the direct evaluation of the triangle diagrams.
$C_1$ is then combined
with the expectation values of the correlators
formed by the corresponding operator
in order to get a prediction for the 
gluonic decay rate of the Higgs boson.
Finally, the conclusions are presented in Section~\ref{seccon}.


\section{Theoretical framework}

Let us in this section fix the notation and present the theoretical framework
used for the calculation.
The leading electroweak corrections are conveniently expressed
in terms of the variable
\begin{eqnarray}
x_t &=& \frac{G_F m_t^2}{8\pi^2\sqrt{2}}\,,
\end{eqnarray}
where $m_t$ is the $\overline{\rm MS}$ definition of the top quark mass
which will be used throughout the paper.
Corrections proportional to $x_t$ arise if in addition to the pure QCD
Lagrangian also the couplings of the Higgs boson ($h$) and
the neutral ($\chi$)
and charged ($\phi^\pm$) Goldstone boson to the top
quark are considered.

For the evaluation of the decoupling constants
it is necessary to know also the corresponding renormalization
constants for the coupling and the masses
\begin{eqnarray}
g_s^0\,\,=\,\,\mu^{\varepsilon}Z_gg_s\,, 
\qquad
m_q^0\,\,=\,\,Z_{m_q} m_q\,,
\label{eqren}
\end{eqnarray}
up to the considered order.
Here and in the following $Z_X$ denotes the renormalization constant
in the $\overline{\rm MS}$ scheme. $Z_g$ and $Z_{m_q}$
will be computed in Section~\ref{secren}
together with the renormalization constants defined through
\begin{eqnarray}
&&
\xi^0-1=Z_3(\xi-1)\,,
\qquad
\psi_q^{L,0}\,\,=\,\,\sqrt{Z^L_{2_q}}\psi^L_q\,,
\qquad
\psi_q^{R,0}\,\,=\,\,\sqrt{Z^R_{2_q}}\psi^R_q\,,
\nonumber\\&&
G_\mu^{0,a}\,\,=\,\,\sqrt{Z_3}G_\mu^a\,,
\qquad
c^{0,a}\,\,=\,\,\sqrt{\tilde{Z}_3}c^a\,,
\label{eqren2}
\end{eqnarray}
up to ${\cal O}(\alpha_s^2 x_t)$.
$g_s=\sqrt{4\pi\alpha_s}$ is the QCD gauge coupling, $\mu$ is the
renormalization scale and $D=4-2\varepsilon$ is the dimensionality of space
time. $m_q$ is the $\overline{\rm MS}$ mass of the light quark masses.
$G_\mu^a$ is the gluon field, and
$c^a$ is the Faddeev-Popov-ghost field.
Colour indices for quark fields, $\psi^{L/R}_q$, are suppressed 
for simplicity. However, it is necessary to distinguish between the
renormalization mode of the right and the left handed quark field,
$\psi^{L/R}=(1\pm\gamma^5)\psi/2$
as they are treated differently in the electroweak theory.
For QCD we have, of course, $Z^L_{2_q}=Z^R_{2_q}$.
The gauge parameter, $\xi$, is defined through the gluon propagator in lowest
order,
\begin{equation}
\frac{i}{q^2+i\epsilon}\left(-g^{\mu\nu}+\xi\frac{q^\mu q^\nu}{q^2}\right).
\label{eqcov}
\end{equation}
The index ``0'' marks the bare quantities. Starting from the three-loop
order, ${\cal O}(\alpha_s^2 x_t)$, the renormalization constant for the
fermion wave function and
mass, $Z_{2_q}^{L/R}$ and $Z_{m_q}$,
depend on the quark species which is indicated by
the additional index $q$. It represents one of the flavours
$u, d, s, c$ or $b$.
Eqs.~(\ref{eqren}) and~(\ref{eqren2}) also hold for $q=t$.
If we refer to the four lightest quarks only the index $q$ will be replaced
by $l$.

In analogy it is possible to write down the relations between
the quantities in the effective (marked by a prime)
and the full theory. Thereby we
restrict ourselves to the case of the coupling constant and
quark masses:
\begin{eqnarray}
g_s^{0\prime}\,\,=\,\,\zeta_g^0 g_s^0\,,\qquad
&
m_q^{0\prime}\,\,=\,\,\zeta_{m_q}^0 m_q^0\,.\qquad
\label{eqdec}
\end{eqnarray}  
Here, of course, $q$ represents only one of the quarks $u,d,s,c$ or $b$.
It is more convenient to consider in a first step bare quantities and 
to perform the renormalization afterwards arriving at:
\begin{eqnarray}
\alpha_s^\prime(\mu)
& = & 
\left(\frac{Z_g}{Z_g^\prime}\zeta_g^0\right)^2\alpha_s(\mu)
\,\,=\,\,
\zeta_g^2\alpha_s(\mu),
\label{eqdecg}\\
m_q^\prime(\mu)
& = &
\frac{Z_{m_q}}{Z_{m}^\prime}\zeta_{m_q}^0 m_q(\mu) 
\,\,=\,\,
\zeta_{m_q}m_q(\mu).
\label{eqdecm}
\end{eqnarray}
In the order considered in this paper the quantities in the effective
theory are independent of the quark species. Thus, 
for the primed quantities the additional ``q'' is absent.
A detailed derivation of the formulae for the computation of
$\zeta_g^0$ and $\zeta_{m_q}^0$ was presented in~\cite{CheKniSte98dec}.
In turned out that only those diagrams where at least one top quark line
is present have to be considered. They have to be evaluated for
vanishing external momentum.
The quite compact formulae for $\zeta_g^0$ and $\zeta_{m_q}^0$ read:
\begin{eqnarray}
\zeta_g^0 &=& \frac{1+\Gamma_{G\bar{c}c}^{0h}(0,0)}
  {\left(1+\Pi_c^{0h}(0)\right)\sqrt{1+\Pi_G^{0h}(0)}}
\,,
\\
\zeta_{m_q}^0 &=& \frac{1-\Sigma_S^{0h}(0)}{1+\Sigma_V^{0h}(0)}
\,,
\end{eqnarray}
where $\Sigma_V(p^2)$ and $\Sigma_S(p^2)$ are the vector and
scalar components
of the light-quark self-energy, defined through
$\Sigma(p)={\not\!p}\left(\Sigma_{V}(p^2)+\gamma^5\Sigma_A(p^2)\right)
  +m_q\Sigma_S(p^2)$.
Note, however, that the axial-vector part $\Sigma_A(p^2)$ does
not enter explicitly into our analysis of $\zeta_{m_q}$ as only leading order
corrections in $x_t$ are considered~\cite{CheKniSte97hbb}.
The dependence of $\Sigma(p)$ on ``q'' is suppressed.
$\Pi_G(p^2)$ and $\Pi_c(p^2)$ are the gluon and ghost vacuum polarizations
and $\Gamma_{G\bar{c}c}(p,q)$ is obtained from the one-particle irreducible
diagrams contributing to the $G\bar{c}c$ green
function~\cite{CheKniSte98dec}.
The superscript ``h'' indicates that only the hard part of the
respective quantities needs to be computed, i.e. only the diagrams involving
the heavy quark contribute.
The remaining paper is concerned with corrections of ${\cal O}(x_t)$.
Therefore in the following the full theory still contains the
top quark, i.e. $n_f=6$ is the number of active flavours,
and in the effective one the top quark is integrated out ($n_f=5$).
Note that in contrast to the renormalization constants of
Eqs.~(\ref{eqren}) and~(\ref{eqren2})
the decoupling constants also receive contributions
from the finite part of the loop integrals.

In this paper the hadronic decay of a scalar Higgs boson in the so-called
intermediate-mass range, i.e. $M_H\aalt M_W$ is considered.
This process is affected by the virtual presence of the heavy top quark.
Therefore it is promising to construct an effective Lagrangian which 
describes the coupling of the Higgs boson to (light) quarks and gluons.
This has already been done in some detail in preceding 
works~\cite{Klu75,InaKubOka83,Spi84,CheKniSte97hbb}.
Hence only a brief sketch of the main steps and
a collection of the relevant formulae is given. The starting point is the
Yukawa Lagrange density describing the coupling of the $H$ boson to quarks,
\begin{eqnarray}
{\cal L}_Y &=& -\frac{H^0}{v^0}
\sum_{q\in\{u,d,s,c,b,t\}}m_{q}^0\bar\psi_{q}^0\psi_{q}^0\,,
\label{eqyuk}
\end{eqnarray}
where the sum runs over all quark flavours. In the limit $m_t\to\infty$
Eq.~(\ref{eqyuk}) can be written as a sum over five 
operators~\cite{Klu75,Spi84} formed by light degrees of freedom
accompanied by coefficient functions containing the residual dependence
on the top quark:
\begin{eqnarray}
{\cal L}_{\rm eff} &=& -\frac{H^0}{v^0}\sum_{i=1}^5C_i^0{\cal O}_i^\prime\,.
\label{eqeff}
\end{eqnarray}
It turns out that only two of the operators, in the following called
${\cal O}_1^\prime$ and ${\cal O}_2^\prime$, contribute to physical processes.
Expressed in terms of bare fields they read:
\begin{eqnarray}
{\cal O}^\prime_1
\,\,=\,\,
\left(G^{0\prime,a}_{\mu\nu}\right)^2\,,
&\quad&
{\cal O}^\prime_2
\,\,=\,\,
\sum_{q\in\{u,d,s,c,b\}}m_{q}^{0\prime}
  \bar\psi_{q}^{0\prime}\psi_{q}^{0\prime}\,.
\end{eqnarray}
The renormalized versions of ${\cal O}_1^\prime$ and
${\cal O}_2^\prime$ and the corresponding coefficient functions
are given by:
\begin{eqnarray}
\left[{\cal O}_1^\prime\right]
&=&
\left[1+2\left(\frac{\alpha_s^\prime\partial}{\partial\alpha_s^\prime}
\ln Z_g^\prime\right)\right]{\cal O}_1^\prime
-4\left(\frac{\alpha_s^\prime\partial}{\partial\alpha_s^\prime}
\ln Z_m^\prime\right){\cal O}_2^\prime\,,
\nonumber\\
\left[{\cal O}_2^\prime\right]
&=&
{\cal O}_2^\prime\,,
\\
C_1
&=&
\frac{1}{1+2(\alpha_s^\prime\partial/\partial\alpha_s^\prime)
  \ln Z_g^\prime}C_1^0\,,
\nonumber\\
C_{2_q}
&=&
\frac{4(\alpha_s^\prime\partial/\partial\alpha_s^\prime)\ln Z_m^\prime}
  {1+2(\alpha_s^\prime\partial/\partial\alpha_s^\prime)\ln Z_g^\prime}C_1^0
  +C_{2_q}^0\,\,.
\label{eqc1c2ren}
\end{eqnarray}
Of course, the leading $m_t$ corrections considered in this paper only
influence the coefficient functions as the operators are defined
in the effective theory where no top quark is present.

The factor $H^0/v^0$ receives a finite universal renormalization 
and can be written in the form
\begin{eqnarray}
\frac{H^0}{v^0} &=& 2^{1/4}G_F^{1/2}H\left(1+\bar{\delta}_u\right)\,.
\label{eqdelu}
\end{eqnarray}
In~\cite{KniSte95} $\bar{\delta}_u$ was evaluated up to
${\cal O}(\alpha_s^2 x_t)$. In our analysis only the ${\cal O}(\alpha_s x_t)$
terms enter. They are given by~\cite{KwiSte94KniSpi94}:
\begin{eqnarray}
\bar{\delta}_u &=& x_t\left[
  \frac{7}{2}
+ \frac{\alpha_s^{(6)}(\mu)}{\pi}\left(
    \frac{19}{3}
  - 2\zeta(2)
  + 7 \ln\frac{\mu^2}{m_t^2}
  \right)
\right]
\,.
\end{eqnarray}
$\zeta$ is Riemann's zeta function, with the 
value $\zeta(2)=\pi^2/6$.

\begin{figure}[t]
 \begin{center}
 \begin{tabular}{c}
   \leavevmode
   \epsfxsize=15.5cm
   \epsffile[110 630 500 730]{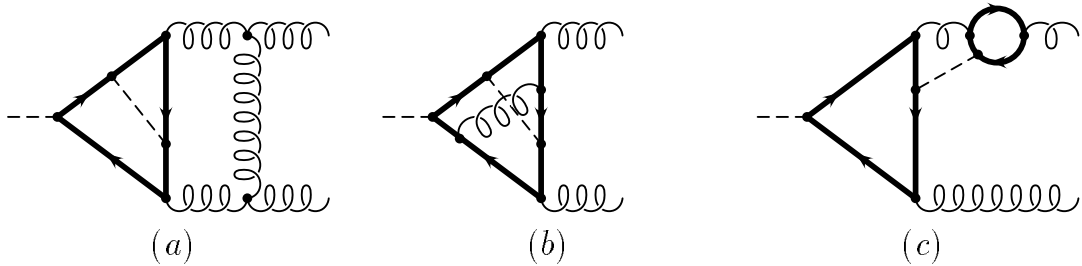}
 \end{tabular}
\caption{\label{fighgg}
  Feynman diagrams contribution to the coefficient function $C_1$.
  The internal dashed line either represents the Higgs boson ($h$)
  or the neutral ($\chi$) or
  charged ($\phi^\pm$) Goldstone boson. The external dashed line corresponds
  to the Higgs boson.
}
 \end{center}
\end{figure}

Once the effective Lagrangian is at hand the evaluation
of the hadronic decay rate splits into two parts, namely the
computation of the imaginary part of the vacuum expectation
values formed by the operators and the calculation of the coefficient
functions. The correlators can be taken over form earlier
works~\cite{CheKniSte97hgg,Che97CheSte97}
as only pure QCD corrections are involved.
In~\cite{CheKniSte97hgg,CheKniSte98dec}
two methods have been used for the computation
of $C_1$. The first one relies on the direct evaluation
of the triangle diagrams (see Fig.~\ref{fighgg}) in the limit of a heavy 
top quark. An expansion in the external momenta has to be performed
up to linear order and the transversal structure has to be projected out.
The second method is based on a low-energy theorem (LET) relating the
coefficient functions to the decoupling constants.
In~\cite{CheKniSte98dec} the following
compact formulae were derived
\begin{eqnarray}
C_1 \,\,=\,\, -\frac{1}{2} \frac{m_t^2\partial}{\partial m_t^2} \ln\zeta_g^2\,,
&\quad&
C_{2_q} \,\,=\,\, 1 + 2 \frac{m_t^2\partial}{\partial m_t^2} \ln\zeta_{m_q}\,,
\label{eqlet}
\end{eqnarray}
which allow for a powerful check.
Concerning the $x_t$ corrections it should be mentioned that the
derivatives do not act on the overall factor $m_t^2$.
In this paper we use both methods in order to compute $C_1$.
Concerning $C_{2_q}$, we use Eq.~(\ref{eqlet}) in order to reproduce the
results given in~\cite{CheKniSte97hbb}.

\section{Renormalization group functions up to ${\cal O}(\alpha_s^2 x_t)$}
\label{secren}

In this section the renormalization constants for different
parameters and fields of the QCD Lagrangian are computed
in the $\overline{\rm MS}$ scheme using dimensional regularization.
Besides QCD corrections also the leading electroweak terms 
proportional to $m_t^2$ are taken into account.
Thus, in principle an index ``$(6)$''
indicating the number of active flavours
would be necessary which is,
however, omitted in this section.
As only pole parts in $D-4$ have to be evaluated it is possible to
reduce the calculation to massless propagator type diagrams,
where the scale is given by the external momentum.
In the case of pure QCD such a strategy is common practice, but
also for the corrections of ${\cal O}(x_t)$ this procedure
is applicable. Even in the presence of a heavy top quark
the needful factors of $m_t$ either arise form the coupling
of the scalar particles to the top quark or from the 
expansion of the internal top quark propagators 
at most up to linear order.

The definition of the renormalization constants is given in 
Eqs.~(\ref{eqren}) and~(\ref{eqren2}).
In the following the computation of 
$Z_3$, $\tilde{Z}_3$, $\tilde{Z}_1$, $Z_{2_q}^{L/R}$
and $Z_{m_q}$ is presented.
The renormalization constants both for the quark-gluon,
the three- and four-gluon vertex can be obtained by the use
of Slavnov-Taylor identities which relate them to the above five
constants~\cite{Muta}:
\begin{eqnarray}
Z_1\,\,=\,\,\frac{Z_3\tilde{Z}_1}{\tilde{Z}_3}\,,
\qquad
Z_{1F_q}\,\,=\,\,\frac{Z_{2_q}\tilde{Z}_1}{\tilde{Z}_3}\,,
\qquad
Z_4\,\,=\,\,\frac{Z_3\tilde{Z}_1^2}{\tilde{Z}_3^2}\,.
\end{eqnarray}
For convenience we list in the following also the pure QCD
results up to order $\alpha_s^2$. They were first
obtained in~\cite{Zas2l}.
Note that from the results for the renormalization constants the
corresponding anomalous dimensions can be computed.

\begin{figure}[t]
 \begin{center}
 \begin{tabular}{c}
   \leavevmode
   \epsfxsize=15.5cm
   \epsffile[110 640 490 730]{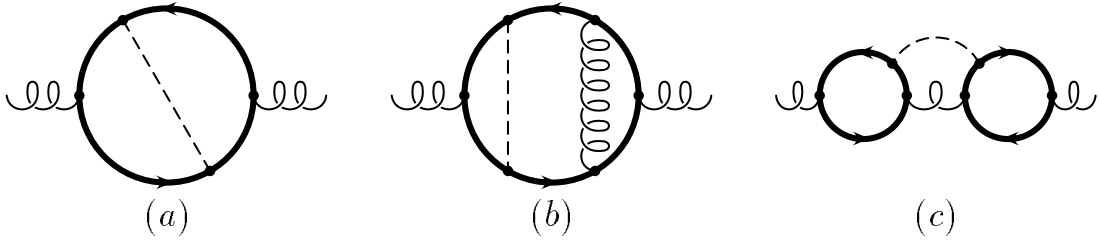}
 \end{tabular}
\caption{\label{figZ3}
  Feynman diagrams contribution to $Z_3$ and
  $\zeta_g^0$. The
  dashed line either represents the Higgs boson ($h$) or the neutral
  ($\chi$) or
  charged ($\phi^\pm$) Goldstone boson.
}
 \end{center}
\end{figure}

From Eqs.~(\ref{eqren2}) it can be seen that for the computation of
$Z_3$ corrections to the gluon propagator have to be considered.
Some sample diagrams contributing at ${\cal O}(\alpha_s^2 x_t)$
are pictured in Fig.~\ref{figZ3} where the dashed lines represents
either the $h$, $\chi$ or $\phi^\pm$ boson. The electroweak corrections
arise for the first time at two loops.
At three-loop order already $224$ diagrams contribute.
$Z_3$ is obtained from the recursive solution of the 
equation
\begin{eqnarray}
Z_3 &=& 1 - K_\varepsilon\left(\Pi_G(q^2) Z_3\right)\,,
\label{eqZ3gen}
\end{eqnarray}
where $\Pi_G$ is the transversal part of the gluon polarization
function defined through
\begin{eqnarray}
\Pi_G^{\mu\nu}(q) &=& \left(-g^{\mu\nu} q^2 + q^\mu q^\nu\right)\Pi_G(q^2)\,.
\end{eqnarray}
The operator $K_\varepsilon$ extracts the pole parts in $\varepsilon$.

After projecting out $\Pi_G(q^2)$ one ends up with purely massless diagrams.
In this case no expansion in $m_t$ is required as the factor $m_t^2$
is provided by the coupling of the scalar bosons to the top quark.
For the generation of the diagrams QGRAF~\cite{Nog93} is used. The output
is then transformed to the package MINCER~\cite{mincer} which is
written in FORM~\cite{FORM} and can deal with one-, two- and three-loop
massless diagrams with one external momentum different from zero,
so-called propagator-type diagrams. After the renormalization of the
parameters $\alpha_s, m_t$ and $\xi$ is performed
for which only renormalization
constants of lower order are needed 
Eq.~(\ref{eqZ3gen}) has to be solved recursively and one gets
\begin{eqnarray}
Z_3 &=& 1 +
  \frac{\alpha_s^{(6)}}{\pi}
  \frac{1}{\varepsilon} \left[
       C_A\,\left(
         \frac{5}{12}
       + \frac{1}{8}\xi
       \right)
       -T\left(n_l+1\right)
         \frac{1}{3}
  +x_t\,T
    \right]
+
  \left(\frac{\alpha_s^{(6)}}{\pi}\right)^2\Bigg\{
\nonumber\\&&\mbox{}
  \frac{1}{\varepsilon^2} \left[
       C_A^2\,\left(
       - \frac{25}{192}
       + \frac{5}{384}\xi
       + \frac{1}{64}\xi^2
       \right)
       +C_A T\left(n_l+1\right)\left(
         \frac{5}{48}
       - \frac{1}{24}\xi
       \right)
    \right]
\nonumber\\&&\mbox{}
  +
  \frac{1}{\varepsilon} \left[
       C_A^2\,\left(
         \frac{23}{128}
       + \frac{15}{256}\xi
       - \frac{1}{128}\xi^2
       \right)
       -C_F T\left(n_l+1\right)
         \frac{1}{8}
       -C_A T\left(n_l+1\right)
         \frac{5}{32}
    \right]
\nonumber\\&&\mbox{}
  +x_t\,T \left[
    \frac{1}{\varepsilon^2} \left(
       - C_F \frac{1}{2}
       + C_A\left(
         - \frac{1}{4}
         + \frac{1}{8}\xi
       \right)
      \right)
    +
    \frac{1}{\varepsilon} \left(
         C_F \frac{1}{4}
       + C_A \frac{25}{48}
      \right)
    \right]
  \Bigg\}
\,,
\end{eqnarray}
where $\alpha_s^{(6)}=\alpha_s^{(6)}(\mu)$.
The superscript ``(6)'' indicates the number of active flavours.
$C_A=N_c$ and $C_F=(N_c^2-1)/(2N_c)$ 
are the Casimir operators of the adjoint and fundamental
representation, respectively, and $T=1/2$ is the trace normalization of the
fundamental representation. $n_l=n_f-1=5$ is the number of light
(massless) quark flavours.

Special care has to be taken for the diagram pictured in 
Fig.~\ref{figZ3}($c$).
If the dashed line corresponds to the $\chi$ boson in each fermion line
exactly one $\gamma_5$ matrix shows up and the naive treatment would fail.
In turns out that the diagram is finite and thus gives no contribution
to $Z_3$. In the next section, however, an analogue diagram contributes to
$\zeta_g^0$ and a careful treatment is mandatory.

In analogy to Eq.~(\ref{eqZ3gen}) the renormalization constant for the
ghost field is obtained form
\begin{eqnarray}
\tilde{Z}_3 &=& 1 - K_\varepsilon\left(\Pi_c(q^2) \tilde{Z}_3\right)
\,.
\label{eqtilZ3gen}
\end{eqnarray}
Corrections of order $x_t$ arise for the first time at three-loop level.
The result reads:
\begin{eqnarray}
\tilde{Z}_3 &=& 1 +
  \frac{\alpha_s^{(6)}}{\pi}
  \frac{1}{\varepsilon}\,C_A\,\left(
        \frac{1}{8}
      + \frac{1}{16}\xi
    \right)
+
  \left(\frac{\alpha_s^{(6)}}{\pi}\right)^2\Bigg\{
  \frac{1}{\varepsilon^2} \left[
       C_A^2\,\left(
       - \frac{1}{16}
       - \frac{3}{256}\xi
       + \frac{3}{512}\xi^2
       \right)
\right.\nonumber\\&&\left.\mbox{}
       +C_AT\left(n_l+1\right)
         \frac{1}{32}
    \right]
  +
  \frac{1}{\varepsilon} \left[
       C_A^2\,\left(
           \frac{49}{768}
         - \frac{1}{512}\xi
       \right)
       -C_A T\left(n_l+1\right)
          \frac{5}{192}
    \right]
\nonumber\\&&\mbox{}
  +x_t\,C_AT\,\left[
    -\frac{1}{8\varepsilon^2}
    +
    \frac{23}{96\varepsilon}
    \right]
  \Bigg\}
\,.
\end{eqnarray}

The renormalization constant $\tilde{Z}_1$ requires the computation of
gluon-ghost vertex diagrams, $\Gamma_{G\bar{c}c}(q,p)$.
For simplicity we choose the gluon
momentum to be zero and again end up with massless propagator-type
integrals. This could in principle introduce unwanted infrared
divergences. However, in~\cite{DavOslTar98} it was shown that this is not
the case and thus $\tilde{Z}_1$ can be computed with the help of
the formula
\begin{eqnarray}
\tilde{Z}_1 &=& 1 
- K_\varepsilon\left(\Gamma_{G\bar{c}c}(q,0)\tilde{Z}_1\right)
\,.
\label{eqtilZ1gen}
\end{eqnarray}
Also for $\tilde{Z}_1$ the $x_t$ corrections in principle appear for the
first time at three-loop order. A closer look to the two-loop
result shows, however, that those diagrams containing a closed fermion loop
add up to zero. As the ${\cal O}(\alpha_s^2 x_t)$ terms come from the
diagrams which contain a top quark loop accompanied with an additional
exchange of a scalar particle we expect that $\tilde{Z}_1$ gets no
${\cal O}(\alpha_s^2 x_t)$ corrections at all. This is verified by an
explicit calculation. For convenience we list the pure QCD result
as it is needed for the computation of $Z_g$:
\begin{eqnarray}
\tilde{Z}_1 &=& 1 +
  \frac{\alpha_s^{(6)}}{\pi}
  \frac{1}{\varepsilon}\,C_A\, \left(
      - \frac{1}{8}
      + \frac{1}{8}\xi
    \right)
+
  \left(\frac{\alpha_s^{(6)}}{\pi}\right)^2\,C_A^2\,\Bigg\{
  \frac{1}{\varepsilon^2}\left(
         \frac{5}{128}
       - \frac{7}{128}\xi
       + \frac{1}{64}\xi^2
    \right)
\nonumber\\&&\mbox{}
  +
  \frac{1}{\varepsilon} \left(
       - \frac{3}{128}
       + \frac{7}{256}\xi
       - \frac{1}{256}\xi^2
    \right)
  \Bigg\}
\,.
\end{eqnarray}

The charge renormalization constant, $Z_g$, can be computed from the
combination of $Z_3, \tilde{Z}_3$ and $\tilde{Z}_1$ with the result
\begin{eqnarray}
Z_g &=& \frac{\tilde{Z}_1}{\tilde{Z}_3\sqrt{Z_3}}
\nonumber\\
    &=& 1 + 
  \frac{\alpha_s^{(6)}}{\pi}
  \frac{1}{\varepsilon} \left[
       -C_A\frac{11}{24}
       +T\left(n_l+1\right)\frac{1}{6}
  -x_t\,T \frac{1}{2}
    \right]
+
  \left(\frac{\alpha_s^{(6)}}{\pi}\right)^2\Bigg\{
\nonumber\\&&\mbox{}
  \frac{1}{\varepsilon^2} \left[
         C_A^2 \frac{121}{384}
       - C_AT\left(n_l+1\right)
          \frac{11}{48}
       + T^2\left(n_l+1\right)^2
          \frac{1}{24}
    \right]
\nonumber\\&&\mbox{}
  +
  \frac{1}{\varepsilon} \left[
       -C_A^2\frac{17}{96}
       +C_A T\left(n_l+1\right)
         \frac{5}{48}
       +C_F T\left(n_l+1\right)
         \frac{1}{16}
    \right]
\nonumber\\&&\mbox{}
  +x_t\,T \left[
    \frac{1}{\varepsilon^2} \left(
         C_A\frac{11}{16}
       + C_F\frac{1}{4}
       - T\left(n_l+1\right)
          \frac{1}{4}
      \right)
    +
    \frac{1}{\varepsilon} \left(
      - C_A\frac{1}{2}
      - C_F\frac{1}{8}
      \right)
    \right]
  \Bigg\}
\,.
\end{eqnarray}
As expected the $\xi$ dependence drops out which is an important
check of the calculation.

From the fermion propagator the wave function renormalization 
constants $Z^{L/R}_{2_q}$
and the one for the mass, $Z_{m_q}$, can be computed. 
Due to the different coupling of the $\chi$ boson
to up- and down-type quarks, one gets an explicit dependence
on the considered quark flavour which is indicated by an additional index.
Although in Section~\ref{secdec} only $Z_{m_q}$ for light quarks is
needed we consider for completeness also $Z_{m_t}$ as the
basic technique of the computation is very similar.
The quark two-point function consists of three parts
\begin{eqnarray}
\Sigma(p) &=&
{\not\!p}\left[
  \Sigma_V(p^2)
 +\gamma_5\Sigma_A(p^2)
\right]
+m_q\Sigma_S(p^2)
\,,
\end{eqnarray}
where $\Sigma_V(p^2)$ and $\Sigma_A(p^2)$
correspond to the vector and axial-vector
and $\Sigma_S(p^2)$ to the scalar part.
Note that in the Standard Model the pseudo-scalar
contribution to $\Sigma(p)$ is zero.
The renormalized fermion propagator can then be written
as~\cite{Aok}
\begin{eqnarray}
S_F^{-1}(p) &=&
{\not\!p}\left[
  \frac{1-\gamma_5}{2} Z_{2_q}^L \left(1+\Sigma_L(p^2)\right)
+ \frac{1+\gamma_5}{2} Z_{2_q}^R \left(1+\Sigma_R(p^2)\right)
\right]
\nonumber\\&&
\quad
-m_q\sqrt{Z^L_{2_q} Z^R_{2_q}}Z_{m_q}\left(1-\Sigma_S(p^2)\right)
\,,
\label{eqferm}
\end{eqnarray}
where $\Sigma_L(p^2)$ and $\Sigma_R(p^2)$ are given by
\begin{eqnarray}
\Sigma_L(p^2)\,\,=\,\,\Sigma_V(p^2) - \Sigma_A(p^2)
\,,\qquad
\Sigma_R(p^2)\,\,=\,\,\Sigma_V(p^2) + \Sigma_A(p^2)
\,.
\end{eqnarray}
From Eq.~(\ref{eqferm}) the following equations can be derived:
\begin{eqnarray}
Z^L_{2_q}&=&1 - K_\varepsilon\left(\Sigma_L(p^2)Z^L_{2_q}\right)
\,,
\nonumber\\
Z^R_{2_q}&=&1 - K_\varepsilon\left(\Sigma_R(p^2)Z^R_{2_q}\right)
\,,
\nonumber\\
\sqrt{Z^L_{2_q} Z^R_{2_q}}Z_{m_q}&=&
1 + K_\varepsilon\left(\Sigma_S(p^2)\sqrt{Z^L_{2_q} Z^R_{2_q}}Z_{m_q}\right)
\,.
\label{eqZLRm}
\end{eqnarray}
Their recursive solutions determine the wave function and mass 
renormalization constants.

\begin{figure}[t]
 \begin{center}
 \begin{tabular}{c}
   \leavevmode
   \epsfxsize=15.5cm
   \epsffile[77 570 560 780]{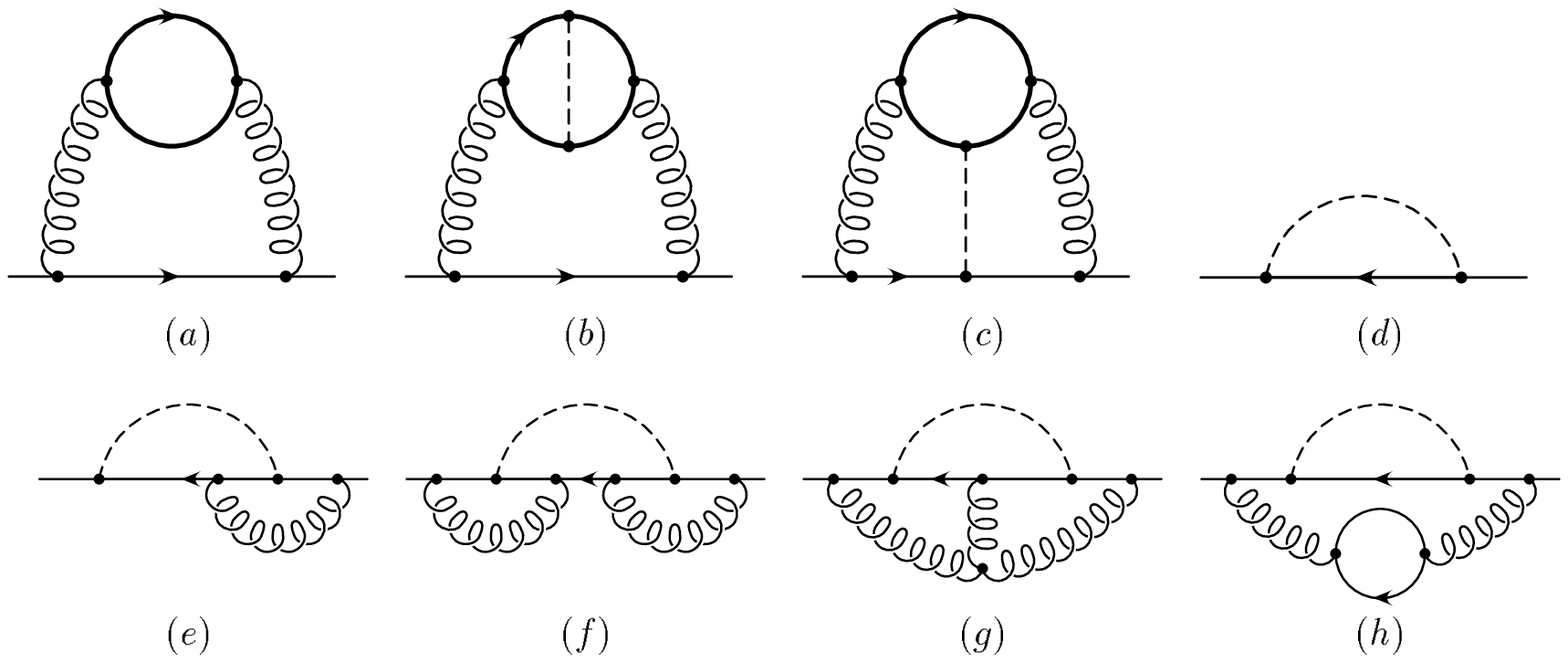}
 \end{tabular}
\caption{\label{figsig}
  Feynman diagrams contribution to $Z_{2_q}$, $Z_{m_q}$ and
  $\zeta_{m_q}^0$. The
  dashed line either represents the Higgs boson ($h$) or the neutral ($\chi$)
  or charged ($\phi^\pm$) Goldstone boson.
}
 \end{center}
\end{figure}

For the computation of $Z^{L/R}_{2_q}$ all masses appearing in the
propagators may be set to zero
and the factors $m_t$ are provided by the Yukawa couplings.
In the case of the four light quarks
only the diagrams where the gluons couple to the considered quark 
contribute
and therefore the renormalization is left-right symmetric:
\begin{eqnarray}
Z^L_{2_l} &=& Z^R_{2_l}
\nonumber\\&=&
1 +
  \frac{\alpha_s^{(6)}}{\pi}
  \frac{1}{\varepsilon}\,C_F\,\left(
         \frac{1}{4}\xi
       - \frac{1}{4}
    \right)
+
  \left(\frac{\alpha_s^{(6)}}{\pi}\right)^2\Bigg\{
  \frac{1}{\varepsilon^2} \left[
         C_F^2\,\left(
         \frac{1}{32}
       - \frac{1}{16} \xi
       + \frac{1}{32} \xi^2
       \right)
\right.\nonumber\\&&\left.\mbox{}
       + C_A C_F\,\left(
         \frac{1}{16}
       - \frac{5}{64} \xi
       + \frac{1}{64} \xi^2
       \right)
    \right]
  +
  \frac{1}{\varepsilon} \left[
         C_F^2 \frac{3}{64}
       + C_A C_F\,\left(
       - \frac{17}{64}
\right.\right.\nonumber\\&&\left.\left.\mbox{}
       + \frac{5}{64} \xi
       - \frac{1}{128} \xi^2
       \right)
       +C_F T\left(n_l+1\right)
         \frac{1}{16}
    \right]
  -x_t\,C_F T
     \frac{1}{4\varepsilon}
  \Bigg\}
\,.
\end{eqnarray}
Next to the pure QCD corrections the class of diagrams pictured in
Fig.~\ref{figsig}($b$) give rise to the ${\cal O}(\alpha_s^2 x_t)$
corrections.
In our approximation the bottom quark is effectively considered to be
massless which has the consequence that the bottom-specific
corrections only contribute to $Z^L_{2_b}$:
\begin{eqnarray}
Z^R_{2_b} &=& Z^R_{2_l}
\,,
\nonumber\\
Z^L_{2_b} &=& Z^L_{2_l} + x_t\Bigg\{
  -\frac{1}{\varepsilon}
+
  \frac{\alpha_s^{(6)}}{\pi}\,C_F\,\left[
  \frac{1}{\varepsilon^2} \left(
         1
       - \frac{1}{4}\xi
    \right)
  +
  \frac{1}{2\varepsilon} 
  \right]
+
  \left(\frac{\alpha_s^{(6)}}{\pi}\right)^2\Bigg\{
\nonumber\\&&\mbox{}
  \frac{1}{\varepsilon^3} \left[
       C_F^2\,\left(
       - \frac{19}{32}
       + \frac{1}{4}\xi
       - \frac{1}{32}\xi^2
       \right)
       + C_A C_F\,\left(
       - \frac{7}{24}
       + \frac{5}{64}\xi
       - \frac{1}{64}\xi^2
       \right)
\right.\nonumber\\&&\left.\mbox{}
       + C_F T\,\left(n_l+1\right)
         \frac{1}{12}
    \right]
  +
  \frac{1}{\varepsilon^2} \left[
       C_F^2\,\left(
       - \frac{23}{64}
       + \frac{1}{8}\xi
       \right)
       + C_A C_F\,\left(
         \frac{151}{192}
       - \frac{5}{64}\xi
       + \frac{1}{128}\xi^2
       \right)
\right.\nonumber\\&&\left.\mbox{}
       - C_F T\,\left(n_l+1\right)
          \frac{7}{48}
     \right]
  +
  \frac{1}{\varepsilon} \left[
       C_F^2\,\left(
         \frac{17}{64}
       - \frac{1}{4}\zeta(3)
       \right)
       + C_A C_F\,\left(
       - \frac{31}{192}
       + \frac{5}{8}\zeta(3)
       \right)
\right.\nonumber\\&&\left.\mbox{}
       + C_F T\,\left(n_l+1\right)
        \frac{1}{16}
     \right]
  \Bigg\}
\Bigg\}
\,,
\end{eqnarray}
with $\zeta(3)\approx1.202\,057$.
For the top quark altogether more diagrams have to be taken into
account as --- in contrast to the bottom case ---
already at one- and two-loop level the exchange of
a Higgs and neutral Goldstone boson may occur.
This is also the reason that both $Z^L_{2_t}$ and $Z^R_{2_t}$
get contributions from the top-specific diagrams. We get:
\begin{eqnarray}
Z^L_{2_t} &=& Z^L_{2_b}
\label{eqZ2tL}
\,,
\end{eqnarray}
and
\begin{eqnarray}
Z^R_{2_t} &=& Z^R_{2_l} + x_t\Bigg\{
  - \frac{2}{\varepsilon}
+
  \frac{\alpha_s^{(6)}}{\pi}\,C_F\,\left[
  \frac{1}{\varepsilon^2} \left(
         2
       - \frac{1}{2}\xi
    \right)
  +
  \frac{1}{\varepsilon}
  \right]
+
  \left(\frac{\alpha_s^{(6)}}{\pi}\right)^2\Bigg\{
\nonumber\\&&\mbox{}
  \frac{1}{\varepsilon^3} \left[
       C_F^2\,\left(
         - \frac{19}{16}
         + \frac{1}{2}\xi
         - \frac{1}{16}\xi^2
       \right)
       + C_A C_F\,\left(
         - \frac{7}{12}
         + \frac{5}{32}\xi
         - \frac{1}{32}\xi^2
       \right)
       + C_F T\,\left(n_l+1\right)
          \frac{1}{6}
    \right]
\nonumber\\&&\mbox{}
  +
  \frac{1}{\varepsilon^2} \left[
       C_F^2\,\left(
         - \frac{23}{32}
         + \frac{1}{4}\xi
       \right)
       + C_A C_F\,\left(
           \frac{151}{96}
         - \frac{5}{32}\xi
         + \frac{1}{64}\xi^2
       \right)
       - C_F T\,\left(n_l+1\right)
          \frac{7}{24}
    \right]
\nonumber\\&&\mbox{}
  +
  \frac{1}{\varepsilon} \left[
       C_F^2\,\left(
            \frac{17}{32}
          - \frac{1}{2}\zeta(3)
       \right)
       + C_A C_F\,\left(
          - \frac{31}{96}
          + \frac{5}{4}\zeta(3)
       \right)
       + C_F T\,\left(n_l+1\right)
           \frac{1}{8}
    \right]
  \Bigg\}
\Bigg\}
\label{eqZ2tR}
\,.
\end{eqnarray}
Note that the top-specific corrections of Eq.~(\ref{eqZ2tR})
are exactly twice as large as the ones of Eq.~(\ref{eqZ2tL}).

As can be seen form Eq.~(\ref{eqZLRm})
the scalar part $\Sigma_S$ can be used together with
$Z^L_{2_q}$ and $Z^R_{2_q}$ in order to compute $Z_{m_q}$.
For it an expansion of $\Sigma(p)$ in $m_q$ up to linear order is
necessary in order to be able to project out $\Sigma_S$ and end up
with massless two-point functions. Notice that here the
factor $m_t$ may also origin from internal top quark propagators.
The class of diagrams pictured in Fig.~\ref{figsig}$(c)$ only contributes
to $\Sigma_S$. Special care has to be taken when the $\chi$ boson is 
exchanged between the top quark loop and the light fermion line as
then each fermion line contains exactly one $\gamma_5$. These diagrams,
however, only develop an overall divergence giving rise to an 
simple $1/\varepsilon$ pole. Therefore we are allowed to adopt a
prescription for $\gamma_5$ according to
't~Hooft and Veltman~\cite{tHoVel73BreiMai77}
which was also used in~\cite{CheKniSte97hbb} and 
which is described in more detail in the next section.
The result for $q\not=b,t$ reads:
\begin{eqnarray}
Z_{m_l} &=&
1 +
  \frac{\alpha_s^{(6)}}{\pi}
  \frac{1}{\varepsilon}\,C_F
       \left(-\frac{3}{4}\right)
+
  \left(\frac{\alpha_s^{(6)}}{\pi}\right)^2\Bigg\{
  \frac{1}{\varepsilon^2} \left(
          C_F^2 \frac{9}{32}
        +  C_A C_F\frac{11}{32}
        - C_F T \left(n_l+1\right) \frac{1}{8}
    \right)
\nonumber\\&&\mbox{}
  +
  \frac{1}{\varepsilon} \left(
        - C_F^2 \frac{3}{64}
        - C_A C_F \frac{97}{192}
        + C_F T \left(n_l+1\right) \frac{5}{48} 
    \right)
\nonumber\\&&\mbox{}
  +x_t\,C_F T \left[
    \frac{1}{2\varepsilon^2}
    +
    \frac{1}{\varepsilon} \left(
         \frac{13}{24}
       - 3\zeta(3)
       + 2 I_{3l}\left(
          \frac{1}{2}
        + 3\zeta(3)
       \right)
      \right)
  \right]
  \Bigg\}
\,.
\end{eqnarray}
The ${\cal O}(\alpha_s^2x_t)$ corrections arise from the diagrams shown
in Fig.~\ref{figsig}$(b)$ and $(c)$.
$I_{3l}$ is the third component of the weak isospin, i.e. $I_{3l}=+1/2$
for up-type quarks and $I_{3l}=-1/2$ for down-type quark flavours.
The case of the bottom quark exhibits more structures and gives the
result
\begin{eqnarray}
Z_{m_b} &=& Z_{m_d} + x_t\Bigg\{
  -\frac{3}{2\varepsilon}
+
  \frac{\alpha_s^{(6)}}{\pi}\,C_F\,\left[
  \frac{9}{4\varepsilon^2} 
  -
  \frac{3}{2\varepsilon} 
  \right]
+
  \left(\frac{\alpha_s^{(6)}}{\pi}\right)^2\Bigg\{
\nonumber\\&&\mbox{}
  \frac{1}{\varepsilon^3} \left[
        - C_F^2 \frac{117}{64}
        - C_A C_F \frac{55}{64}
        + C_F T\left(n_l+1\right) \frac{5}{16}
    \right]
  +
  \frac{1}{\varepsilon^2} \left(
          C_F^2\frac{261}{128}
        + C_A C_F\frac{285}{128}
\right.\nonumber\\&&\mbox{}\left.
        - C_F T\left(n_l+1\right)\frac{17}{32}
    \right)
  +
  \frac{1}{\varepsilon} \left(
       C_F^2\,\left(
        - \frac{61}{128}
        + \frac{33}{8}\zeta(3)
       \right)
       + C_A C_F\,\left(
        - \frac{805}{384}
        - \frac{15}{16} \zeta(3)
       \right)
\right.\nonumber\\&&\mbox{}\left.
       + C_F T\,\left(n_l+1\right)
         \frac{41}{96}
    \right)
  \Bigg\}
\Bigg\}
\,.
\end{eqnarray}
Finally for the case of the top quark we obtain
\begin{eqnarray}
Z_{m_t} &=& Z_{m_u} + x_t\Bigg\{
  \frac{3}{2\varepsilon}
+
  \frac{\alpha_s^{(6)}}{\pi}\,C_F\,\left[
  -\frac{9}{4\varepsilon^2} 
  +
  \frac{3}{2\varepsilon}
  \right]
+
  \left(\frac{\alpha_s^{(6)}}{\pi}\right)^2\Bigg\{
\nonumber\\&&\mbox{}
  \frac{1}{\varepsilon^3} \left[
          C_F^2 \frac{117}{64}
        + C_A C_F \frac{55}{64}
        - C_F T\left(n_l+1\right) \frac{5}{16}
    \right]
  +
  \frac{1}{\varepsilon^2} \left(
        - C_F^2\frac{261}{128}
        - C_A C_F\frac{285}{128}
\right.\nonumber\\&&\mbox{}\left.
        + C_F T\left(n_l+1\right)\frac{17}{32}
    \right)
  +
  \frac{1}{\varepsilon} \left(
       C_F^2\,\left(
          \frac{157}{128}
        - \frac{9}{8}\zeta(3)
       \right)
       + C_A C_F\,\left(
          \frac{239}{128}
        - \frac{9}{16} \zeta(3)
       \right)
\right.\nonumber\\&&\mbox{}\left.
       - C_F T\,\left(n_l+1\right)
         \frac{11}{32}
    \right)
  \Bigg\}
\Bigg\}
\,.
\end{eqnarray}
It is remarkable that except for the $1/\varepsilon$ pole at
${\cal O}(\alpha_s^2 x_t)$
the coefficients of the structures
$x_t(\alpha_s^{(6)}/\pi)^n/\varepsilon^m$ coincide in the
expressions for $Z_{m_b}$ and $Z_{m_t}$ up to an overall sign.


\section{Decoupling relations}
\label{secdec}

The computation of the decoupling constant relating $\alpha_s$
in the full and the effective theory requires essentially
three ingredients:
the hard part of the gluon polarization function, the one for the
ghost polarization function and the one of the ghost-gluon vertex.
At the one-loop order altogether only one diagram 
contributes  to $\Pi_G^{0h}$, namely the one containing a 
closed top quark loop which is obviously
gauge invariant. This is also the case for the ${\cal O}(\alpha_s x_t)$
result where inside the top loop an additional scalar particle is exchanged.
At two-loop level $\Pi_c^{0h}$ and $\Gamma_{G\bar{c}c}^{0h}$ only
receive pure QCD contributions. Actually, the individual
diagrams contributing to $\Gamma_{G\bar{c}c}^{0h}$ give
non-vanishing contributions, the sum, however, adds up to zero.
At three-loop level there are 224, 14, respectively, 98 diagrams
which have to be taken into account in order to compute
of $\Pi_G^{0h}$, $\Pi_c^{0h}$, respectively, $\Gamma_{G\bar{c}c}^{0h}$
up to order $\alpha_s^2 x_t$.
Again, the separate diagrams contributing to $\Gamma_{G\bar{c}c}^{0h}$
add up to zero.
$\Pi_c^{0h}$ gets non-vanishing corrections which have to
be combined with $\Pi_G^{0h}$.
For the generation of the diagrams the program QGRAF~\cite{Nog93}
is used. The output is then transformed into a format suitable for the
package MATAD~\cite{Ste96} which is written in FORM~\cite{FORM}
for the purpose to compute one-, two- and three-loop tadpole diagrams.

A special treatment is necessary for the class of diagrams pictured in
Fig.~\ref{figZ3}($c$)
where the dashed line represents the neutral CP odd Goldstone boson, $\chi$.
Here, $\gamma^5$ occurs in two different fermion lines and the naive
treatment would lead to a wrong result.
Instead we follow the work of 
't~Hooft and Veltman~\cite{tHoVel73BreiMai77}
and write $\gamma^5$ in the form
\begin{eqnarray}
\gamma^5 &=& \frac{i}{4!}\epsilon_{\mu\nu\rho\sigma}
             \gamma^{[\mu\nu\rho\sigma]}\,,
\label{eqgamma5}
\end{eqnarray}
where $\gamma^{[\mu\nu\rho\sigma]}$ is the anti-symmetrized product
of four $\gamma$ matrices.
The $\epsilon$ tensor is pulled off from the analytical expression
and an object containing eight indices is obtained:
\begin{eqnarray}
\Pi^{\mu^\prime\nu^\prime\rho^\prime\sigma^\prime}_{G,\mu\nu\rho\sigma}(q^2)
&=&
\Pi_{G,1}(q^2)q^2g_{[\mu}^{[\mu^\prime}g_\nu^{\nu^\prime}
g_\rho^{\rho^\prime}g_{\sigma]}^{\sigma^\prime]}
+\Pi_{G,2}(q^2)q_{[\mu}q^{[\mu^\prime}g_\nu^{\nu^\prime}
g_\rho^{\rho^\prime}g_{\sigma]}^{\sigma^\prime]}\,.
\label{eqpig}
\end{eqnarray}
$\Pi_{G,1}$ and $\Pi_{G,2}$ are functions of $q^2$ and may be
extracted with the help of the projectors~\cite{CheKniSteBar98}:
\begin{eqnarray}
P_{1,\mu\nu\rho\sigma}^{\mu^\prime\nu^\prime\rho^\prime\sigma^\prime}(q)
&=&\frac{24}{(q^2)^2}\,\frac{\left(
q^2g_{[\mu}^{[\mu^\prime}g_\nu^{\nu^\prime}g_\rho^{\rho^\prime}
g_{\sigma]}^{\sigma^\prime]}
-4q_{[\mu}q^{[\mu^\prime}g_\nu^{\nu^\prime}g_\rho^{\rho^\prime}
g_{\sigma]}^{\sigma^\prime]}\right)}{(D-1)(D-2)(D-3)(D-4)}\,,
\nonumber\\
P_{2,\mu\nu\rho\sigma}^{\mu^\prime\nu^\prime\rho^\prime\sigma^\prime}(q)
&=&\frac{96}{(q^2)^2}\,\frac{\left(
-q^2g_{[\mu}^{[\mu^\prime}g_\nu^{\nu^\prime}g_\rho^{\rho^\prime}
g_{\sigma]}^{\sigma^\prime]}
+Dq_{[\mu}q^{[\mu^\prime}g_\nu^{\nu^\prime}g_\rho^{\rho^\prime}
g_{\sigma]}^{\sigma^\prime]}\right)}{(D-1)(D-2)(D-3)(D-4)}\,.
\label{pro}
\end{eqnarray}
Note that both $P_1$ and $P_2$ develop a $1/\varepsilon$ pole which is
a consequence of the fact that for $D=4$ both structures appearing
in Eq.~(\ref{eqpig}) are linear dependent.
This artificial pole, however, cancels in each diagram individually
leading to a finite result. This was actually expected as
there is no contribution to $Z_g$ from this class of
diagrams at the order considered in this paper. 

Some comments concerning the renormalization are in order.
The parameters in the lower order diagrams have to be replaced
by the corresponding renormalized values. Concerning
pure QCD, the counterterms for $\alpha_s$ and $m_t$ have to be known
up to order $\alpha_s x_t$. In principle also the QCD gauge parameter
$\xi$ appears in the individual contributions
$\Pi_G^{0h}$, $\Pi_c^{0h}$ and $\Gamma_{G\bar{c}c}^{0h}$, however,
only starting from three loops, which has no effect on the
terms of order $\alpha_s^2 x_t$.
The parameters which are present in the diagrams contributing to
the ${\cal O}(\alpha_s x_t)$ results
receive only contributions from pure QCD counterterms.

In order to compute the renormalized quantity $\zeta_g$ also
the renormalization constants $Z_g^{(5)}$ and $Z_g^{(6)}$ are needed.
We choose to express the r.h.s. of Eq.~(\ref{eqdec}) in terms
of $\alpha_s^{(6)}$. Therefore $Z_g^{(5)}$ gets a dependence
on $m_t$ through the substitution of $\alpha_s^{(5)}$
whereas before only pure QCD terms were present.
$Z_g^{(6)}$ has to be known up to order 
$\alpha_s^2 x_t$ which was derived in the previous section.
Finally we obtain for $\zeta_g$ the following result:
\begin{eqnarray}
\zeta_g^2 &=& 1 +
  \frac{\alpha_s^{(6)}(\mu)}{\pi}\,T\,\Bigg\{
       - \frac{1}{3} \ln\frac{\mu^2}{m_t^2}
+
  \frac{\alpha_s^{(6)}(\mu)}{\pi} \left[
    C_F\,\left(
       - \frac{13}{48}
       + \frac{1}{4} \ln\frac{\mu^2}{m_t^2} 
    \right)
\right.\nonumber\\&&\left.\mbox{}
    +C_A\,\left(
         \frac{2}{9}
       - \frac{5}{12} \ln\frac{\mu^2}{m_t^2}
    \right)
       + T\,\frac{1}{9} \ln^2\frac{\mu^2}{m_t^2}
  \right]
+ x_t\Bigg\{
       - \frac{2}{3}
       + \ln\frac{\mu^2}{m_t^2} 
\nonumber\\&&\mbox{}
+
  \frac{\alpha_s^{(6)}(\mu)}{\pi} \left[
     C_F\,\left(
       - \frac{17}{16}
       + \frac{5}{4}\zeta(2)
       + \frac{25}{8}\zeta(3)
       - 3 \ln\frac{\mu^2}{m_t^2} 
       + \frac{3}{4} \ln^2\frac{\mu^2}{m_t^2}
     \right)
\right.\nonumber\\&&\left.\mbox{}
     +C_A\,\left(
       - \frac{5}{4} 
       + \frac{3}{8}\zeta(2)
       - \frac{95}{64}\zeta(3)
       + \frac{7}{4} \ln\frac{\mu^2}{m_t^2} 
     \right)
     +T\,\left( 
         \frac{5}{4}
       + \frac{7}{8}\zeta(3)
       + \frac{4}{9} \ln\frac{\mu^2}{m_t^2} 
\right.\right.\nonumber\\&&\left.\left.\mbox{}
       - \frac{2}{3} \ln^2\frac{\mu^2}{m_t^2} 
     \right) 
       - \frac{7}{2}\zeta(3) T 
    \right]
  \Bigg\}
\Bigg\}
\label{eqzetag}
\,,
\end{eqnarray}
where the contribution of the diagrams in  
Fig.~\ref{figZ3}($c$) corresponds to the last entry
in the last line of Eq.~(\ref{eqzetag}). 
For convenience also the pure QCD result of ${\cal O}(\alpha_s^2)$
is listed. The corresponding three-loop terms can be found 
in~\cite{CheKniSte97als,CheKniSte98dec}.

The decoupling constants for the light masses, $\zeta_{m_q}$,
requires the computation of the hard part of the fermion propagator.
Implicitly this has already been done in~\cite{CheKniSte97hbb}
where, however, the main focus was on the evaluation of $C_{2_q}$, the
effective coupling to light quarks. In this paper
we will also list the result for $\zeta_{m_q}$.
As already mentioned above $\zeta_{m_q}$
depends on the considered quark flavour which can be seen by a look to
the diagrams pictured in Fig.~\ref{figsig}.
Some words are in order in connections with the diagram in
Fig.~\ref{figsig}$(c)$ when the dashed line corresponds to the $\chi$ boson.
Actually this diagram is responsible for the difference between
$\zeta_{m_u}$ and $\zeta_{m_d}$, as the $\chi\bar{f}f$ coupling
is proportional to the third component of the isospin.
Furthermore, the treatment of $\gamma_5$ needs some care.
As already mentioned in Section~\ref{secren} the diagrams of this class
have an overall divergence which is reflected in the $1/\varepsilon$ pole
contributing to $Z_{m_q}$ whereas all subdiagrams are finite.
Thus we are allowed to adopt the prescription for $\gamma_5$ described above.
The decoupling constant for the $u, d, s$ and $c$ quark then reads
(for simplicity we set $N_c=3$):
\begin{eqnarray}
\zeta_{m_l} &=& 1
+
  \left(\frac{\alpha_s^{(6)}(\mu)}{\pi}\right)^2\Bigg\{
         \frac{89}{432} 
       - \frac{5}{36}\ln\frac{\mu^2}{m_t^2} 
       + \frac{1}{12}\ln^2\frac{\mu^2}{m_t^2}
\nonumber\\&&\mbox{}
+ x_t\left[
         \frac{101}{144} 
       - \frac{5}{12}\zeta(2) 
       + \frac{73}{12}\zeta(3) 
       - 9\zeta(4) 
       - \frac{7}{6}\ln\frac{\mu^2}{m_t^2}
       + 6\zeta(3)\ln\frac{\mu^2}{m_t^2} 
\right.\nonumber\\&&\mbox{}\left.
       + 2I_{3l}\left(
         - \frac{37}{18} 
         - \frac{19}{3}\zeta(3) 
         + 9\zeta(4) 
         - \ln\frac{\mu^2}{m_t^2} 
         - 6\zeta(3)\ln\frac{\mu^2}{m_t^2} 
       \right)
  \right]
\Bigg\}
\,.
\label{eqzetaml}
\end{eqnarray}
For the bottom quark one receives:
\begin{eqnarray}
\zeta_{m_b} &=& \zeta_{m_d} + x_t\Bigg\{
    \frac{5}{4} 
  + \frac{3}{2}\ln\frac{\mu^2}{m_t^2}
+
  \frac{\alpha_s^{(6)}(\mu)}{\pi} \left[
      \frac{16}{3} 
    - 4\zeta(2) 
    + \frac{7}{2}\ln\frac{\mu^2}{m_t^2} 
    + \frac{3}{2}\ln^2\frac{\mu^2}{m_t^2}
  \right]
\nonumber\\&&\mbox{}
+
  \left(\frac{\alpha_s^{(6)}(\mu)}{\pi}\right)^2 \left[
      \frac{472933}{12096} 
    - \frac{6133}{168}\zeta(2) 
    + \frac{905}{72}\zeta(3) 
    + \frac{383}{18}\zeta(4) 
    + \frac{1251}{112}S_2 
    + \frac{19}{72}D_3
\right.\nonumber\\&&\mbox{}\left.
    - \frac{7}{9}B_4
    + \left(
        \frac{763}{18}
      - \frac{55}{3}\zeta(2)
      - \frac{43}{4}\zeta(3)
    \right)\ln\frac{\mu^2}{m_t^2}
    + \frac{529}{48}\ln^2\frac{\mu^2}{m_t^2} 
    + \frac{29}{12}\ln^3\frac{\mu^2}{m_t^2} 
\right.\nonumber\\&&\mbox{}\left.
    + n_l\left(
      - \frac{23}{24}
      + \frac{31}{36}\zeta(2)
      - 2\zeta(3)
      + \left(
        - \frac{241}{144}
        + \frac{2}{3}\zeta(2)
      \right)\ln\frac{\mu^2}{m_t^2}
      - \frac{1}{2}\ln^2\frac{\mu^2}{m_t^2}
\right.\right.\nonumber\\&&\mbox{}\left.\left.
      - \frac{1}{12}\ln^3\frac{\mu^2}{m_t^2}
    \right)
  \right]
\Bigg\}
\,,
\label{eqzetamb}
\end{eqnarray}
where we have used $C_F=4/3$, $C_A=3$ and $T=1/2$.
The constants
\begin{eqnarray}
S_2&=&{4\over9\sqrt3}\mbox{Cl}_2\left({\pi\over3}\right)
\,\,\approx\,\,0.260\,434
\,,
\nonumber\\
D_3&=&6\zeta(3)-\frac{15}{4}\zeta(4)
     -6\left(\mbox{Cl}_2\left({\pi\over3}\right)\right)^2
\,\,\approx\,\,-3.027\,009
\,,
\nonumber\\
B_4&=&16\li\left({1\over2}\right)-{13\over2}\zeta(4)-4\zeta(2)\ln^22
+{2\over3}\ln^42
\,\,\approx\,\,-1.762\,800
\,,
\end{eqnarray}
where $\zeta(4)=\pi^4/90$, $\mbox{Cl}_2$ is Clausen's function and 
$\mbox{Li}_4$ is the quadrilogarithm,
occur in the evaluation of the three-loop master
diagrams~\cite{Bro92,Avd95Che95,Bro98}.
In~\cite{CheKniSte98dec} the three-loop corrections of
${\cal O}(\alpha_s^3)$ were computed.
The results of Eqs.~(\ref{eqzetaml}) and~(\ref{eqzetamb}) will be used
in the next section in order to compute $C_{2_l}$ and $C_{2_b}$.


\section{Hadronic Higgs decay}
\label{sechad}

In this section we compute the coefficient function $C_1$ using two
different methods. The first one is concerned with
the direct evaluation of the triangle diagrams (see Fig.~\ref{fighgg})
connecting the Higgs boson to two gluons. In a second step $C_1$
is evaluated with the help of the LET where the result of the decoupling
constant $\zeta_g$ derived in the previous section is used.

At three-loop level altogether $990$~diagrams contribute to the
order $\alpha_s^2 x_t$. Some typical examples
are pictured in Fig.~\ref{fighgg} where the internal dashed line either
represents the Higgs boson, $h$, the neutral Goldstone boson, $\chi$,
or the charged Goldstone boson, $\phi^\pm$. In the latter case
also the bottom quark is present in the fermion loop. The external
Higgs boson, however, only couples to top quarks as corrections
proportional to $x_t$ are considered.
The diagrams have to be expanded in both external momenta and the transversal
structure which arises from the external gluons is projected out in order
to end up with scalar integrals.

Again, the packages QGRAF~\cite{Nog93} and MATAD~\cite{Ste96} 
are used for the generation, respectively, the computation
of the diagrams. A general gauge parameter, $\xi$, for 
QCD\footnote{In the considered limit the electroweak gauge parameters
  drop out trivially.} is used and the independence of the
final result serves as a welcome check for the correctness of the
result.
There is again a class of diagrams which requires a special treatment
due to the fact that $\gamma_5$ appears in two different
fermion lines. A sample diagram is shown in Fig.~\ref{fighgg}$(c)$.
Of course, this class is tightly connected to the one discussed
in connection with $\zeta_g$ (see Fig.~\ref{figZ3}$(c)$) and
we can adopt the handling for $\gamma_5$ developed in Section~\ref{secdec}.
From Eq.~(\ref{eqzetag}) one can see that no $\ln m_t$ terms
are present which has the consequence that according to the 
low-energy theorem of Eq.~(\ref{eqlet}) $C_1$ gets no contribution
from these diagrams. This is confirmed by the direct calculation of
the vertex diagrams: there is no contribution from this class
of graphs.

After taking into account the counterterms needed to get the renormalized
coefficient function (see Eq.(\ref{eqc1c2ren})) one arrives at:
\begin{eqnarray}
C_1 &=& -\frac{1}{6}\,T\,\frac{\alpha_s^{(6)}(\mu)}{\pi}\Bigg\{
  1 
  - 3 x_t
+ \frac{\alpha_s^{(6)}(\mu)}{\pi}\left[
  - C_F\,\frac{3}{4}
  + C_A\,\frac{5}{4}
  - T\,\frac{1}{3} \ln\frac{\mu^2}{m_t^2} 
\right.\nonumber\\&&\left.\mbox{}
  + x_t\left(
    C_F\,\left(
        9
      - \frac{9}{2} \ln\frac{\mu^2}{m_t^2} 
    \right)
    - C_A\,\frac{21}{4}
    +T\,\left(
      - \frac{2}{3} 
      + 2 \ln\frac{\mu^2}{m_t^2}
    \right)
  \right)
\right]
\Bigg\}
\,.
\label{eqC1}
\end{eqnarray}
The ${\cal O}(\alpha_sx_t)$ terms can be found in~\cite{DjoGam94} and the
${\cal O}(\alpha_s^2)$ results were 
computed in~\cite{InaKubOka83,DjoSpiZer91}.
The QCD corrections of ${\cal O}(\alpha_s^3)$ and ${\cal O}(\alpha_s^4)$ 
are also known~\cite{CheKniSte97hgg,CheKniSte98dec}, however,
for simplicity they are not displayed in Eq.~(\ref{eqC1}).

If we use the decoupling constant of Eq.~(\ref{eqzetag}) and plug it into
Eq.~(\ref{eqlet}) we obtain the identical result which serves as a
non-trivial check. Note that the diagrams to be considered in
both approaches are quite different.
Furthermore, the method based on the LET requires the renormalization
constant $Z_g$ to be known at ${\cal O}(\alpha_s^2 x_t)$ whereas
for the direct computation only the terms up to order
$\alpha_s x_t$ are necessary.

For completeness we also list the result for $C_{2_q}$ obtained from
Eqs.~(\ref{eqzetaml}), respectively, (\ref{eqzetamb})
and Eq.~(\ref{eqlet}). For the light quarks we get
\begin{eqnarray}
C_{2_l} &=& 1+
\left(\frac{\alpha_s^{(6)}(\mu)}{\pi}\right)^2 \left[
    \frac{5}{18} 
  - \frac{1}{3}\ln\frac{\mu^2}{m_t^2}
  + x_t \left(  \frac{7}{3} 
        - 12\zeta(3) 
        + 2I_{3l}\left(
            2
          + 12\zeta(3)
        \right)
  \right)
\right]
\,.
\end{eqnarray}
The coefficient function for the bottom quark reads:
\begin{eqnarray}
C_{2_b} &=& C_{2_d} + x_t\Bigg\{
    - 3
+
  \frac{\alpha_s^{(6)}(\mu)}{\pi} \left[
    - 7 
    - 6\ln\frac{\mu^2}{m_t^2}
  \right]
+ 
\left(\frac{\alpha_s^{(6)}(\mu)}{\pi}\right)^2 \left[
     - \frac{12169}{144} 
     + \frac{110}{3}\zeta(2) 
\right.\nonumber\\&&\left.\mbox{}
     + \frac{43}{2}\zeta(3)
     - \frac{89}{2}\ln\frac{\mu^2}{m_t^2} 
     - \frac{55}{4}\ln^2\frac{\mu^2}{m_t^2} 
     + n_l\left(
         \frac{241}{72}
       - \frac{4}{3}\zeta(2)
       + 2\ln\frac{\mu^2}{m_t^2}
\right.\right.\nonumber\\&&\left.\left.\mbox{}
       + \frac{1}{2}\ln^2\frac{\mu^2}{m_t^2}
     \right)
  \right]
\Bigg\}
\,.
\end{eqnarray} 
In~\cite{CheKniSte97hbb} $C_{2_l}$ and $C_{2_b}$ are listed for general
gauge group $SU(N_c)$.

Let us now have a look at the numerical consequences on the decay rate.
Therefore  also the imaginary part of the correlator
$\langle[{\cal O}_1^\prime][{\cal O}_1^\prime]\rangle$ is needed
which can be found in~\cite{CheKniSte97hgg}.
Furthermore the universal corrections arising from $\bar{\delta}_u$
(see Eq.~(\ref{eqdelu})) have to be
taken into account. Inserting all building blocks into the equation
\begin{eqnarray}
\Gamma\left(H\to gg\right) &=& 
\frac{\sqrt{2}G_F}{M_H}\left(1+\bar{\delta}_u\right)^2
C_1^2\,\mbox{Im}\langle[{\cal O}_1^\prime][{\cal O}_1^\prime]\rangle
\end{eqnarray}
and expanding up to the three-loop level leads to
\begin{eqnarray}
\frac{\Gamma\left(H\to gg\right)}{\Gamma^{\rm Born}(H\to gg)}
&=&
  1 
+ x_t
+ \frac{\alpha_s^{(5)}(M_H)}{\pi}\left[
    23.750
   - 1.667 n_l
\right.\nonumber\\&&\left.\mbox{}
   + x_t\left(
      38.837
     + 2.000 \ln\frac{M_H^2}{m_t^2}
     - 1.167 n_l
   \right)
\right]
+ \left(\frac{\alpha_s^{(5)}(M_H)}{\pi}\right)^2
\nonumber\\&&\mbox{}
\left[
       370.196
     -  47.186 n_l
     +   0.902 n_l^2
     + \left(
          2.375
        + 0.667 n_l
       \right) \ln\frac{M_H^2}{m_t^2}
  \right]
\nonumber\\
&=&
  1 
+ x_t
+ \frac{\alpha_s^{(5)}(M_H)}{\pi}\left[
    17.917
   + x_t\left(
      33.004
     + 2.000 \ln\frac{M_H^2}{m_t^2}
   \right)
\right]
\nonumber\\&&\mbox{}
+ \left(\frac{\alpha_s^{(5)}(M_H)}{\pi}\right)^2\left[
       156.808
     + 5.708 \ln\frac{M_H^2}{m_t^2}
  \right]
\,,
\end{eqnarray}
with
$\Gamma^{\rm Born}(H\to gg)
  =G_FM_H^3/36\pi\sqrt2 \times (\alpha_s^{(5)}(M_H)/\pi)^2$.
The renormalization scale $\mu^2$ is set to $M_H^2$.
The resulting logarithms $\ln M_H^2/m_t^2$ are numerically
small which makes a resummation not
necessary~\cite{CheKniSte97hgg,CheKniSteBar98}.
In a similar way to the leading electroweak corrections~\cite{DjoGam94}
also at ${\cal O}(\alpha_s x_t)$ large cancellations between
the universal and non-universal terms take place.
Actually, the ``1'' in front of the ${\cal O}(x_t)$ term
is composed of $(7-6)$ and
the ``33.004'' results from $(131.504-98.500)$
where in both cases the first number corresponds to the
universal corrections.
Choosing $m_t=m_t(M_H)= 173$~GeV and $M_H=100$~GeV
leads to
\begin{eqnarray}
\frac{\Gamma\left(H\to gg\right)}{\Gamma^{\rm Born}(H\to gg)}
&=&
  1 
+ x_t
+ \frac{\alpha_s^{(5)}(M_H)}{\pi}\left(
     17.917
   + 30.811 x_t
\right)
+ 150.550\left(\frac{\alpha_s^{(5)}(M_H)}{\pi}\right)^2  
\nonumber\\
&\approx&
1 
+ 0.0031 
+ 0.6559 
+ 0.0035
+ 0.2017 
\,.
\end{eqnarray}
If one compares the ${\cal O}(x_t)$ corrections to the
pure QCD terms of the same loop order they are clearly negligible.
The three-loop corrections of ${\cal O}(\alpha_s x_t)$ amount roughly
$1$~\% of order $\alpha_s^2$ terms.
It is, however, interesting to note that they
have the same sign and that they are of the same order of magnitude as the
two-loop corrections of order $x_t$.


\section{Conclusions}
\label{seccon}

In this paper the gluonic decay of a scalar Higgs
boson in the intermediate mass range is considered.
The pure QCD corrections up to ${\cal O}(\alpha_s^4)$
were computed recently~\cite{CheKniSte97hgg}
and it turned out that they are quite sizeable.
On the other hand the leading electroweak corrections are small.
The main focus of the present paper was the evaluation
of additional QCD corrections to the ${\cal O}(x_t)$ term.
Although the new corrections are small as compared to the
QCD terms they are quite sizeable as compared to the
leading electroweak corrections.
It seems that once QCD is switched on
and its full structure is available,
i.e., the gluon self-interaction is at work, large corrections
can be expected.
This is also motivated by the comparison of the leading order
QCD corrections to $\Gamma(H\to gg)$ and
$\Gamma(H\to \gamma\gamma)$.
In the latter case only the Abelian part of the QCD enters and
the correction factor only amounts to 
$1-2$~\%~\cite{ZhengWu90}.

In this paper the top quark is
considered to be much heavier than the other quarks and the
Higgs boson. However, the generalization of the analysis 
to any heavy quark which fulfills these conditions
is straightforward.


\vspace{2em}

\centerline{\large\bf Acknowledgments}
\noindent
I would like to thank K.G. Chetyrkin, P.A. Grassi and B.A. Kniehl
for useful discussions and comments.

\vspace{2em}


\end{document}